\documentclass[prb,aps,twocolumn,showpacs,floats]{revtex4}
\usepackage{epsfig}

\begin{document}

%------------------------------------------------------------------------------
\title{First-principles study of the polar O--terminated ZnO surface\\
       in thermodynamic equilibrium with oxygen and hydrogen}

\author{B. Meyer}
\affiliation{Lehrstuhl f\"ur Theoretische Chemie,
             Ruhr-Universit\"at Bochum, 44780 Bochum, Germany}
\date{\today}
%------------------------------------------------------------------------------

\begin{abstract}
Using density-functional theory in combination with a thermodynamic formalism
we calculate the relative stability of various structural models of the polar
O--terminated (000$\bar{1}$)--O surface of ZnO. Model surfaces with different
concentrations of oxygen vacancies and hydrogen adatoms are considered.
Assuming that the surfaces are in thermodynamic equilibrium with an O$_2$
and H$_2$ gas phase we determine a phase diagram of the lowest-energy
surface structures.
For a wide range of temperatures and pressures we find that hydrogen will
be adsorbed at the surface, preferentially with a coverage of 1/2 monolayer.
At high temperatures and low pressures the hydrogen can be removed and a
structure with 1/4 of the surface oxygen atoms missing becomes the most stable
one. The clean, defect-free surface can only exist in an oxygen-rich
environment with a very low hydrogen partial pressure. However, since we find
that the dissociative adsorption of molecular hydrogen and water (if also the
Zn--terminated surface is present) is energetically very preferable, it is
very unlikely that a clean, defect-free (000$\bar{1}$)--O surface can be
observed in experiment.
\end{abstract}


\maketitle

%------------------------------------------------------------------------------

\section{Introduction}
\label{sec:intro}

To understand the structure, composition and stability of polar surfaces
on a solid theoretical basis is one of the challenging problems in surface
science.\cite{noguera}  The most interesting polar surfaces are so called
``Tasker-type(3)'' surfaces\cite{tasker} which are formed by alternating
layers of oppositely charged ions. Assuming a purely ionic model\cite{comment1}
in which all ions are in their formal bulk oxidation state, such a stacking
sequence creates a dipole moment perpendicular to the surfaces which diverges
with slab thickness, and with simple electrostatic arguments it can be shown
that the surface energy will diverge with sample size.\cite{tasker}  To quench
the dipole moment and to make the polar surfaces stable, a redistribution
of charges in the surface layers has to take place.\cite{nosker}  For various
polar surfaces different mechanisms to accomplish the charge compensation
have been observed,\cite{fks}  however, in many cases the underlying
stabilization mechanism is very controversially discussed in the literature.

One of the most widely investigated examples of Tasker-type(3) polar surfaces
are the two basal planes of ZnO: the O--terminated (000$\bar{1}$)--O and the
Zn--terminated (0001)--Zn surface. The two surfaces are the terminating planes
of a stacking sequence of hexagonal Zn and O layers along the crystallographic
$c$-axis with alternating distances of $R_1$=0.61\,{\AA} and $R_2$=1.99\,{\AA}.
In this case, as can be easily shown\cite{noguera,zno34}, the polar ZnO
surfaces are only stable if the O--terminated face is less negative and the
Zn--terminated surface layer less positively charged compared to the formal
bulk oxidation state by a factor of $R_1/(R_1+R_2)$$\approx$1/4. In principle,
three different scenarios are conceivable to accomplish this charge
redistribution: The ionic charge of the surface ions may be reduced from
$\pm$2 to $\pm$3/2, which may be regarded as an ``electron transfer'' from
the O-- to the Zn--terminated surface (Ia). As a result, partially occupied
surface bands will appear with a 3/4 filled O--$2p$ band at the
(000$\bar{1}$)--O and a 1/4 filled Zn--$4s$ band at the (0001)--Zn
surface. This is often referred to as ``intrinsic surface state
compensation''\cite{nosker} or as ``metalization of the polar
surfaces''.\cite{wander}  However, whether a true metallic state is present
will depend on the dispersion of the partially occupied bands. Additionally,
in a second step, the surface may reconstruct and undergo a distortion in
which, for example for the O--terminated surface, four surface atoms combine
in such a way that an unoccupied $2p$--band splits from the other eleven
occupied $2p$--bands and the surface becomes insulating again (Ib).
Secondly, the charge reduction of the surface layers may take place by
removing 1/4 of the surface ions (II). The so created vacancies may be ordered
and form a reconstruction or may be randomly distributed. Finally, charged
species may be adsorbed to reduce the formal oxidation state of the surface
ions (III). For example, water may dissociate and protons (H$^+$) and hydroxyl
groups (OH$^-$) adsorb on every second O and Zn surface ion,
respectively.\cite{comment2}  All three scenarios represent idealizations of
the charge compensation process. In general, any combinations of the three
mechanisms are conceivable, like the simultaneous formation of vacancies and
partially filled bands, as long as the charge compensation rule is obeyed.
The surface structure finally realized for a specific polar surface will then
be the one with the lowest surface energy.

For ZnO it was believed for a long time that both polar surfaces exist in
an unreconstructed, truncated-bulk-like state. After standard preparation
procedures both surfaces show regular (1$\times$1) pattern in low-energy
electron diffraction (LEED)\cite{duke} and other diffraction
experiments.\cite{wander,jedrecy,overbury}  Some evidence for missing
Zn ions on the (0001)--Zn surface was found in gracing incidence X-ray
diffraction (GIXD)\cite{jedrecy}, however, for the (000$\bar{1}$)--O surface
no evidence for substantial amounts of surface oxygen vacancies was detected
in GIXD\cite{jedrecy} and low-energy alkali-ion scattering
(LEIS).\cite{overbury}

For ideal, truncated-bulk-like surface terminations only mechanism (Ia) can
explain the stability of the polar ZnO surfaces. Consequently, in most
theoretical first-principles studies of the polar ZnO surfaces ideal
surface terminations together with partially filled surface bands were
assumed.\cite{wander,carlsson,zno34}  Studies exploring the other
two stabilizations mechanisms are very scarce. In a pioneering ab-initio
study Wander and Harrison\cite{wander_h} investigated whether the polar
surfaces may be stabilized by the dissociation of water and the adsorption of
H$^+$ and OH$^-$ groups according to mechanism (III). They found this
energetically unfavorable compared to situation (Ia). However, instead of
1/2 monolayers, which would be the ideal configurations for band filling, only
full monolayers of H$^+$ and OH$^-$ were considered, thereby overcompensating
the needed charge transfer between the polar surfaces. In addition, only one
specific adsorption site for the H$^+$ and OH$^-$ groups was probed. 

Meanwhile two recent experiments have created considerable doubt that the
polar ZnO surfaces really exist in a clean, unreconstructed state. With
scanning tunneling microscopy (STM) it was shown\cite{diebold1,diebold2} that
the Zn--terminated surface is characterized by the presence of nanoscaled,
triangular islands with a height of one ZnO double-layer. The shape of the
islands is size-dependent and typical pit structures appear for larger
islands. Since the step edges are O--terminated, the high step concentration
leads to a significant decrease of Zn ions in the surface. A rough analysis
of the island and pit size distribution yielded that approximately 1/4 of
the Zn ions is missing, in agreement with mechanism (II). With detailed
density-functional theory (DFT) calculations\cite{diebold1,kresse} it was
confirmed that a crystal termination with triangular shaped islands and pits
is indeed lower in energy than the perfect bulk-truncated surface for a wide
range of oxygen and hydrogen chemical potentials. Under H rich conditions,
structures with up to 1/2 monolayer of hydroxyl groups were even more stable,
indicating that the actual surface morphology will sensitively depend on the
chemical environment.

On the other hand, for the O-terminated polar surface no such island and pit
structure was found with STM.\cite{diebold2}   However, with He scattering
(HAS) it was discovered\cite{woell} that at ultrahigh vacuum (UHV) conditions
O--terminated surfaces with (1$\times$1) LEED and HAS diffraction pattern are
always hydrogen covered, whereas after a careful removal of the hydrogen a
(1$\times$3) structure is found. The (1$\times$3) spots are best visible in
HAS, but under certain conditions can also be observed in LEED.\cite{woell}
The H--free surface is very reactive and dissociates molecular hydrogen and
water and therefore exits only for a limited time even at UHV conditions.
In a subsequent study\cite{zno35,zno36,staemmler,fink} CO was used as a probe
molecule to distinguish between clean and hydrogen saturated surfaces. By
comparing calculated CO adsorption energies for different surface structures
with experimental results it was confirmed that the polar O--terminated
surface is usually hydrogen covered whereas a clean (1$\times$1)
(000$\bar{1}$)--O surface is very unlikely to exist.

In previous studies the H coverage of the polar O--terminated surface was
not observed most likely because LEED and X--rays are not sensitive to
hydrogen. However, it is not clear how the structures found in the HAS study,
Ref.~\onlinecite{woell}, lead to a stabilization of the polar O--terminated
surface. A full hydrogen monolayer would overcompensate the charge transfer
so that again partially occupied bands have to be present, and the nature of
the H--free (1$\times$3) structure is still unknown. There is some evidence
from X--ray photoelectron spectroscopy (XPS)\cite{woell} that oxygen vacancies
are involved, but 1/3 of the oxygens missing is also not the expected vacancy
concentration to stabilize the surface.

Motivated by these experimental findings we explore in the present paper
the competition between the three stabilization mechanisms in a very general
way. The main focus will be on the O--terminated (000$\bar{1}$)--O
surface, and we take an approach very similar in spirit to the investigation
of the Zn--terminated surface in Ref.~\onlinecite{diebold1} and
\onlinecite{kresse}. For a series of surface models we determine the total
energies and the fully relaxed atomic structures using a first-principles DFT
approach. Surface structures with various oxygen vacancy concentrations and
different amounts of adsorbed hydrogen are considered, including structures
corresponding to the three ideal stabilization scenarios and structures
compatible with the HAS observations.

Static total-energy DFT calculations only give results for zero temperature,
zero pressure and for surfaces in contact with vacuum. However, the actual
lowest-energy structure of the (000$\bar{1}$)--O surface will depend on the
environment and can change with temperature $T$, pressure $p$ and exposure to
O$_2$ and H$_2$ gas phases. Therefore, to determine the equilibrium structure
and composition of the surface at finite temperature and oxygen and hydrogen
partial pressures, we combine our DFT results with a thermodynamic description
of the surfaces. To take deviations in surface composition and the presence of
gas phases into account, we introduce appropriate chemical
potentials\cite{chadi} and calculate an approximation of the Gibbs free
surface energy.\cite{finnis1}  Depending on the chemical potentials we then
determine the surface structure with the lowest free energy which allows us
to construct a phase diagram for the surface. If we assume that the surface
is in thermodynamic equilibrium with the gas phases, we can relate the
chemical potentials to a given temperature $T$ and pressure $p$. In this way
we are able to extend our zero temperature and zero pressure DFT results to
experimentally relevant environments, thereby bridging the gap between
UHV-like conditions and temperatures and gas phase pressures that are
typically applied, for example, in catalytic processes like the methanol
synthesis.\cite{catalysis}

\newpage

%------------------------------------------------------------------------------

\section{Computational Approach}
\label{sec:theorie}

\subsection{Thermodynamics}
\label{sec:therm}

In this section we will give a brief description of the thermodynamic
formalism which we have used to determine the most stable structures of
the polar O--terminated ZnO surface. The formalism has been successfully
applied in several previous surface
studies\cite{padilla,wang,batyrev,pojani,reuter,diebold1} and is described
in more detail in Refs.~\onlinecite{finnis1} and \onlinecite{reuter}.

The general expression for the free energy of a surface in equilibrium with
particle reservoirs at the temperature $T$ and pressure $p$ is given
by\cite{cahn}
\begin{equation}
\label{def_gsurf}
\gamma(T,p) = \frac{1}{A} \left( G(T,p,\{N_i\}) -
                                 \sum_i N_i\, \mu_i(T,p) \right) \;,
\end{equation}
where $G(T,p,\{N_i\})$ is the Gibbs free energy of the solid with the
surface of interest, $A$ is the surface area, and $\mu_i$, $N_i$ are the
chemical potentials and particle numbers of the various species. In contrast
to the usual convention in macroscopic thermodynamics we define here the
chemical potentials per atom rather than per mole. For the study of the polar
O--terminated ZnO surface in contact with an oxygen and a hydrogen gas phase
we have to consider the three chemical species $i$~= Zn, O and H.

For simplicity we have assumed two independent reservoirs for O$_2$ and
H$_2$ with a common pressure $p$. Experimentally, it is more likely that a
mixture of O$_2$ and H$_2$ is present. In this case, the pressure $p$ in
Eq.~(\ref{def_gsurf}) has to be replaced by appropriate partial pressures
$p_{\rm O_2}$ and $p_{\rm H_2}$. However, in the present study we will keep
the restriction of separate reservoirs in the sense that we do not allow O$_2$
and H$_2$ to react to H$_2$O, which would be the case in full thermodynamic
equilibrium. This is justified by arguing that the reaction barrier for the
formation of H$_2$O is high enough that the reaction plays no role on the
time scales of interest.

In thermodynamic equilibrium the chemical potentials would be determined
uniquely by the temperature $T$, the pressure $p$ and the total particle
numbers of the solid and the gas phases. The surface structure, here
represented by the particle numbers $N_i$, would then be determined by a
unconstrained minimization of the surface free energy, Eq.~(\ref{def_gsurf}).
However, this is not very practical to do. Therefore we will take a different
approach: We calculate the surface free energy of a series of model surfaces
with different structures and compositions as a function of the chemical
potentials. For given chemical potentials we predict which surface structure
is the most stable one by searching for the surface model with the lowest
surface free energy. In a second step, the chemical potentials are then
related to actual temperature and pressure conditions by assuming that the
surface is in thermodynamic equilibrium with the gas phases.

In our calculations all surfaces are represented by periodically repeated
slabs so that the Gibbs free energy $G(T,p,\{N_i\})$ refers to the content
of one supercell. Since ZnO is not centrosymmetric, slabs representing
the polar ZnO surfaces are inevitably O--terminated on one side and
Zn--terminated on the other side. It is therefore not possible to assign
unique surface energies to the two polar surface terminations. Only the sum
of the surface energies and thereby the cleavage energy are well defined
quantities. However, in the present study we are only interested in the
{\em relative} stability of O--terminated surfaces with different structures
and compositions. The Zn-face of the slabs is unchanged in all calculations.
Therefore we relate all energies relative to a reference state which we have
taken to be the ideal, truncated-bulk termination. We define the change of the
cleavage energy $\Delta\gamma$ as the difference of Eq.~(\ref{def_gsurf}) for
a slab with a chosen surface structure and a slab with ideal surface
terminations:
\begin{eqnarray}
\label{def_dgamma}
\Delta\gamma(T,p)
 & = & \frac{1}{A} \Big(
       G^{\rm surf}_{\rm slab}(T,p,N_{\rm V},N_{\rm H}) -
       G^{\rm ref}_{\rm slab}(T,p) \nonumber \\
 &   & {}+ N_{\rm V}\,\mu_{\rm O}(T,p) -
       N_{\rm H}\,\mu_{\rm H}(T,p) \Big) \;.
\end{eqnarray}
Here, $G^{\rm surf}_{\rm slab}$ and $G^{\rm ref}_{\rm slab}$ are the Gibbs
free energies of the supercells with the model surface and the reference
configuration, respectively, and $A$ is now the area of the surface
unit cell. Since we only consider structures of the O--terminated surface
with O--vacancies and adsorbed H atoms, only the number of O--vacancies
$N_{\rm V}$ (= difference of the number of O--atoms in the slab with the
model surface and the reference state) and the number of adsorbed H atoms
$N_{\rm H}$ appears in Eq.~(\ref{def_dgamma}), i.e. the chemical potential
of Zn drops out. The difference $\Delta\gamma$ is negative if a model surface
is more stable than the ideal, truncated-bulk-like surface termination and
positive otherwise.

In principle we now have to calculate the Gibbs free energy of all slabs
representing our surface models, including the contributions coming from
changes in volume and in entropy. The entropy term may be calculated, for
example, by evaluating the vibrational spectra in a quasiharmonic
approximation\cite{frank}, but in practice this is computationally very
demanding. As is apparent from Eq.~(\ref{def_dgamma}), only the
{\em difference} of the Gibbs free energy of two surface structures enters the
expression for $\Delta\gamma$. In Ref.~\onlinecite{reuter} it was shown that
the vibrational contributions to the entropy usually cancel to a large extent
and that the influence of volume changes are even smaller. Therefore we will
neglect all entropy and volume effects. The Gibbs free energies then reduce to
the internal energies of the slabs and we can replace $G^{\rm surf}_{\rm slab}$
and $G^{\rm ref}_{\rm slab}$ in Eq.~(\ref{def_dgamma}) by the energies as
directly obtained from total--energy (e.g. DFT) calculations.

Finally we have to determine meaningful ranges in which we can vary the
chemical potentials. First, there are upper bounds for all three chemical
potentials $\mu_{\rm O}$, $\mu_{\rm H}$ and $\mu_{\rm Zn}$, beyond which
molecular oxygen and molecular hydrogen would condensate and metallic Zn
would crystallize at the surface. These bounds are given by the total energy
of the isolated molecules $E_{\rm O_2}$, $E_{\rm H_2}$ and of bulk Zn
$E^{\rm bulk}_{\rm Zn}$ (neglecting volume and entropy effects):
\begin{equation}
\label{upper_bound}
\mu_{\rm O}  \le \frac{1}{2}E_{\rm O_2}\;,\quad
\mu_{\rm H}  \le \frac{1}{2}E_{\rm H_2}\;,\quad
\mu_{\rm Zn} \le E^{\rm bulk}_{\rm Zn}\;.
\end{equation}
In the following we will use these upper bounds as zero point of energy
and relate the chemical potentials relative to the total energies of the
isolated molecules:
\begin{equation}
\label{def_dmu}
\Delta\mu_{\rm O} = \mu_{\rm O} - \frac{1}{2}E_{\rm O_2}\;,\quad
\Delta\mu_{\rm H} = \mu_{\rm H} - \frac{1}{2}E_{\rm H_2}\;.
\end{equation}
Furthermore we impose that the surface is always in equilibrium with the
ZnO bulk phase. Then the sum of $\mu_{\rm O}$ and $\mu_{\rm Zn}$ has to be
equal to the total energy $E_{\rm ZnO}$ of bulk ZnO. Thus only one of the
two chemical potentials $\mu_{\rm O}$ and $\mu_{\rm Zn}$ is independent, and
together with Eq.~(\ref{upper_bound}) we introduce lower bounds for the
chemical potentials. Using the energy of formation $E_{\rm f}$ of ZnO:
\begin{equation}
\label{def_ef}
E_{\rm f} = E_{\rm ZnO} - E_{\rm Zn} - \frac{1}{2}E_{\rm O_2}
\end{equation}
the allowed range for the chemical potential $\Delta\mu_{\rm O}$ is given by:
\begin{equation}
\label{range}
E_{\rm f} \;\le\; \Delta\mu_{\rm O} \;\le\; 0 \;.
\end{equation}
If we assume that the surfaces are in thermodynamic equilibrium with the gas
phases we can relate the chemical potentials $\Delta\mu_{\rm O}$ and
$\Delta\mu_{\rm H}$ to a given temperature $T$ and partial pressures
$p_{\rm O_2}$ and $p_{\rm H_2}$. For ideal gases we can use the well-known
thermodynamic expressions\cite{reuter}
\begin{equation}
\Delta\mu_{\rm O}(T,p_{\rm O_2}) = \frac{1}{2} \Big(
\tilde{\mu}_{\rm O_2}(T,p^0) + k_{\rm B}T \ln(p_{\rm O_2}/p^0) \Big)
\end{equation}
and
\begin{equation}
\Delta\mu_{\rm H}(T,p_{\rm H_2}) = \frac{1}{2} \Big(
\tilde{\mu}_{\rm H_2}(T,p^0) + k_{\rm B}T \ln(p_{\rm H_2}/p^0) \Big)
\end{equation}
in which $p^0$ is the pressure of a reference state and the temperature
dependence of the chemical potentials $\tilde{\mu}_{\rm O_2}(T,p^0)$ and
$\tilde{\mu}_{\rm H_2}(T,p^0)$ is tabulated in thermochemical reference
tables.\cite{janaf}  However, it should be noted, as pointed out by
Finnis\cite{finnis2}, that the equilibrium with the gas phase need not
to be perfect. It is sufficient if the surface is in equilibrium with the
bulk phase. In this case, the chemical potential is related to defect
concentrations of the bulk. For example, if oxygen vacancies are the dominant
defects we have\cite{mayer1,hagen,mayer2}
\begin{equation}
\Delta\mu_{\rm O}(T,p) = \tilde{\mu}_0 - k_{\rm B}T \ln(c_{\rm V}/c_0)
\end{equation}
with the vacancy concentration $c_{\rm V}$ and the oxygen occupancy of
the O--lattice site of $c_0$.

\subsection{Ab-Initio Calculations}
\label{sec:method}

The total energies of slabs representing different model surfaces as 
well as the bulk and molecular reference energies were calculated within
the framework of density-functional theory (DFT).\cite{hks} Exchange and
correlation effects were included in the generalized--gradient
approximation (GGA) using the functional of Perdew, Becke and
Ernzerhof (PBE).\cite{pbe}  Normconserving pseudopotentials\cite{van-pp}
were employed together with a mixed-basis set consisting of plane waves and
non-overlapping localized orbitals for the O--$2p$ and the Zn--$3d$
electrons.\cite{mb}  A plane-wave cut-off energy of 20\,Ry was sufficient
to get well converged results. Monkhorst-Pack k-point meshes\cite{mp}
with a density of at least (6$\times$6$\times$6) points in the primitive ZnO
unit cell were chosen. A dipole correction\cite{bengtsson,bm} to the
electrostatic potential was included in the calculations to eliminate 
all artificial interactions between the periodically repeated supercells
due to the dipole moment of the slabs. For more details on convergence
parameters, the construction of appropriate supercell as well as the
calculated bulk and clean surface structures of ZnO we refer to
Ref.~\onlinecite{zno34}, where the same computational settings as in the
present study were used.

All surfaces were modeled by periodically repeated slabs. Very thick slabs
consisting of 8 Zn-O double-layers were used to reduce the residual internal
electric field.\cite{zno34}. To represent different surface structures
(1$\times$2), (1$\times$3) and (2$\times$2) surface unit cells with different
combinations of O vacancies and H adatoms were considered. All atomic
configurations were fully relaxed by minimizing the atomic forces.

In Table~\ref{tab:ref} we compare the computed binding energies of different
bulk and molecular reference structures with experimental heat of formations.
While the calculated binding energies for isolated $H_2$ molecules and bulk Zn
agree quite well with experiment, there is a noticeable error of 1.2\,eV in
the binding energy of the free O$_2$ molecule. This is a well known deficiency
of DFT.\cite{batyrev,reuter}  The overestimation of the O$_2$ binding energy
is also reflected in a formation energy for ZnO which is to low by 0.6\,eV.
Such deviations would seriously influence our analysis of the surface
energies, Eq.~(\ref{def_dgamma}), and would alter the allowed range for the
oxygen chemical potential, Eq.~(\ref{range}). Therefore, to circumvent errors
introduced by a poor description of the O$_2$ molecule, we have applied the
following procedure: we take the experimental value for the formation energy
$E_{\rm f}$ of ZnO from Table~\ref{tab:ref} and we use Eq.~(\ref{def_ef}) to
replace the total energy of O$_2$ by $E_{\rm f}$ and the total energies of Zn
and ZnO. $E_{\rm Zn}$ and $E_{\rm ZnO}$ are both bulk quantities which are
typically more accurately described in DFT than molecular energies.

\begin{table}[!t]
\noindent
\begin{center}
\begin{minipage}[c]{246pt}
\def\arraystretch{1.5}
\def\tabcolsep{8pt}
\begin{tabular}{lccccc}
  & H$_2$ & O$_2$ & Bulk Zn & Bulk ZnO \\ \hline
$E_{\rm f}^{\rm PBE}$ [eV] & 4.50 & 6.38 & 1.12 & 2.84 \\
$H_{\rm f}^{\rm exp}$ [eV] & 4.52 & 5.17 & 1.35 & 3.50
\end{tabular}
\end{minipage}
\end{center}
\caption{\label{tab:ref}
Calculated binding energies $E_{\rm f}^{\rm PBE}$ for the isolated H$_2$ and
O$_2$ molecules and for the bulk phases of Zn and ZnO. Zero-point vibrations
are not included. The experimental heat of formations $H_{\rm f}^{\rm exp}$
are for $T$=298\,K and $p$=1\,bar and are taken from
Ref.~\protect\onlinecite{nist}.}
\end{table}

%------------------------------------------------------------------------------

\section{Results and Discussion}
\label{sec:results}

\begin{figure}[!t]
\noindent
\epsfxsize=150pt
\epsffile{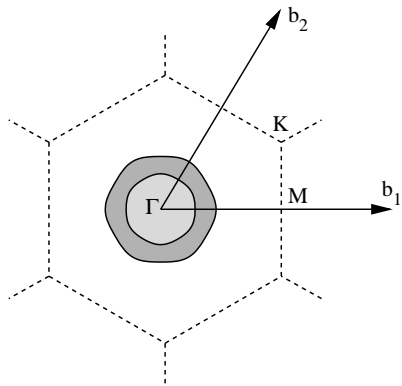}
\caption{\label{fig:fermi}
Contour of the Fermi energy level for the partially occupied
(000$\bar{1}$)--O surface bands plotted within the surface Brillouin zone
(shown to scale). Two O--$2p$--bands cross the Fermi level (thick solid
lines). The unoccupied regions of the two bands are indicated by light gray
and dark gray areas, respectively, ${\bf b}_1$ and ${\bf b}_2$ are the
reciprocal lattice vectors, and the surface Brillouin zone is shown by
dashed lines.}
\end{figure}

\subsection{Surface Distortions}
\label{sec:dist}

First we explore the possibility whether the ideal, truncated-bulk-like
(000$\bar{1}$)--O surface may lower its energy by breaking the symmetry of
the surface layers, thereby adopting a distorted surface structure according
to mechanism (Ib). A tendency for symmetry breaking reconstructions is often
indicated by a strong nesting of the Fermi surface. In Fig.~\ref{fig:fermi}
we have plotted the two-dimensional Fermi surface formed by the partially
occupied O--$2p$ bands. The plot represents a cut through the Brillouin zone
of our supercell including only k-vectors with a zero component perpendicular
to the surface. Figure~\ref{fig:fermi} reveals that actually two surface
bands cross the Fermi level. This is well known and in full agreement with
band structure plots presented in Refs.~\onlinecite{wander} and
\onlinecite{carlsson}.  The corresponding wave functions are strongly localized
at the oxygen atoms of the terminating surface layers and are mainly formed by
the two O--$2p$ orbitals parallel to the surface. By integrating the occupied
and unoccupied areas of the Brillouin-zone we find that indeed roughly 1/2
electron per surface atom is missing to fill the two surface bands. However,
both Fermi contours are almost spherical and only a weak nesting is present.

As a second test, we did several calculations in which we randomly displaced
the surface atoms in the top atomic layer of our slabs and started an atomic
relaxation. Different slabs with (1$\times$2), (1$\times$3) and (2$\times$2)
surface unit cells were used but in all cases the surfaces relaxed back
toward a symmetric structure with a 3-fold symmetry. The surface energy
was always higher than in the fully symmetric state, so that no hint for a
tendency toward symmetry breaking reconstructions was found.

\begin{figure}[!t]
\noindent
\epsfxsize=246pt
\epsffile{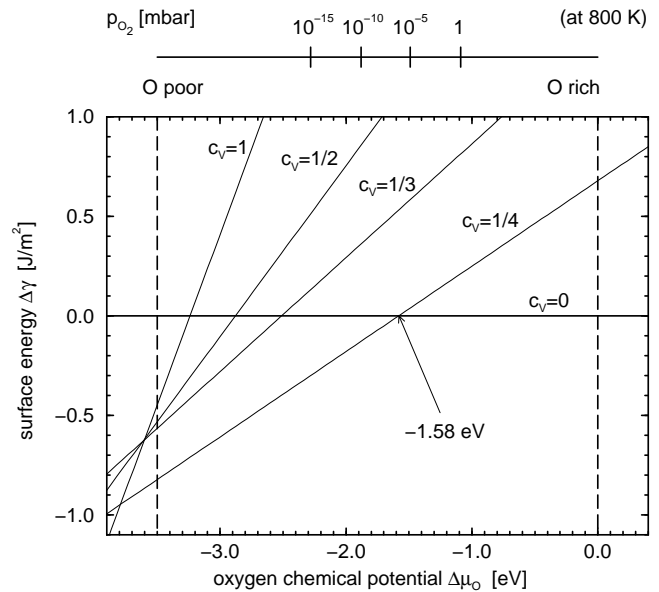}
\caption{\label{fig:ovac}
Surface free energy $\Delta\gamma$ of the polar O-terminated (000$\bar{1}$)--O
surfaces with different concentrations of O-vacancies $c_{\rm V}$ as
function of the oxygen chemical potential $\Delta\mu_{\rm O}$. In the top
$x$--axis, the chemical potential $\Delta\mu_{\rm O}(T,p)$ has been translated
into a pressure scale for the fixed temperature of $T$=800\,K. Vertical dashed
lines indicate the allowed range of $\Delta\mu_{\rm O}$: For higher chemical
potentials O$_2$ will condensate on the surface and for lower values of
$\Delta\mu_{\rm O}$ metallic bulk Zn can form.}
\end{figure}

\subsection{Oxygen Vacancies}
\label{sec:ovac}

In the next step we investigate whether the (000$\bar{1}$)--O surface is
stabilized by creating oxygen vacancies. From slabs with (1$\times$1),
(1$\times$2), (1$\times$3) and (2$\times$2) surface unit cells we have
removed one O--atom, thereby creating vacancy concentrations $c_{\rm V}$
of 1, 1/2, 1/3 and 1/4. In Fig.~\ref{fig:ovac} we have plotted the change
of the surface energy $\Delta\gamma$ of the four defect structures relative
to the defect-free surface as a function of the oxygen chemical potential
$\Delta\mu_{\rm O}$. As to be expected from mechanism (II) the defective
surface with 1/4 of the O atoms missing is the most stable surface structure
over a wide range of chemical potentials. Translating the chemical potential
into temperature and pressure conditions (assuming that the surface is in
equilibrium with an O$_2$ gas phase, see upper $x$--axis in
Fig.~\ref{fig:ovac}) we see that this will be the most stable surface at
typical UHV-conditions of surface science experiments. However, at oxygen
rich conditions (high pressure and low temperature), exceeding a chemical
potential of $-$1.58\,eV, the ideal, defect-free surface becomes the most
stable structure. The other surfaces with 1, 1/2 and 1/3 vacancy
concentrations are higher in energy for all chemical potentials and will
therefore not be present in thermodynamic equilibrium. In particular, it
is very unlikely that the (1$\times$3) structure observed in the HAS
experiment\cite{woell} corresponds to a simple missing-row structure with
every third O atom removed from the surface.

At this point we should emphasize that plots like Fig.~\ref{fig:ovac} strictly
only allow to rule out surface structures which are higher in energy than
other surface models. Since we can only perform calculations for a limited set
of surface models, it is always possible that a not yet considered structure
with a lower energy exists. For example, since we use periodically repeated
surface unit cells of a specific size in our DFT approach, all our defect
structures are perfectly ordered. It is however very well possible, that
disordered or even incommensurate structures might lead to a lower energy.
Additionally, there are hints that island and pit structures like the ones
observed for the Zn--terminated surface\cite{diebold1} may also for the
O--terminated surface be lower in energy than the ideal surface and the
surface with 1/4 vacancy concentration considered in our study.\cite{kresse2}

In Table~\ref{tab:ovac} we have listed O vacancy formation energies
$E_{\rm V}$ which we have defined in the present context as the energy
difference between the ideal surface and a vacancy surface structure of
concentration $c_{\rm V}$ plus $1/2\,E_{\rm O_2}$, i.e.\ $E_{\rm V}$ is
defined with respect to the total energy $E_{\rm O_2}$ of free oxygen
molecules, and not, as usually done, with respect to bulk ZnO. $E_{\rm V}$
depends strongly on the vacancy concentration, indicating a strong interaction
between the vacancies. Up to $c_{\rm V}$=1/4 the energy cost to remove
O--atoms is modest. This is not surprising since up to a vacancy concentration
of 1/4 the removal of O-atoms supports the charge compensation of the
O--terminated surface and will result in a better filling of the partially
occupied O--$2p$ band. For higher vacancy concentrations, however, we start
to overcompensate the charge transfer which stabilizes the surface. The O--$2p$
band is full now, and we have to start to fill Zn--$4s$--states in the
conduction band. Therefore the energy cost to remove more O--atoms increases
rapidly.

\begin{table}[!t]
\noindent
\begin{center}
\begin{minipage}[c]{246pt}
\def\arraystretch{1.5}
\def\tabcolsep{8pt}
\begin{tabular}{lccccc}
$c_{\rm V}$ &  1  &  1/2  &  1/3  & 1/4$^{\rm (a)}$ & 1/4$^{\rm (b)}$\\ \hline
$E_{\rm V}$ [eV] & +3.24 & +2.88 & +2.51 & +1.80 & +1.58
\end{tabular}
\end{minipage}
\end{center}
\caption{\label{tab:ovac}
Calculated vacancy formation energies $E_{\rm V}$ per O atom for removing
oxygen from the ideal surface, forming O$_2$ molecules and a surface structure
with a O vacancy concentration of $c_{\rm V}$. (a) for a 6--fold (2$\times$2)
arrangement of the O vacancies with a separation $2a$, (b) for a rectangular
O vacancy pattern with distances of $2a$ and $\sqrt{3}\,a$, $a$ being the
ZnO lattice constant.}
\end{table}

\begin{table}[!t]
\noindent
\begin{center}
\begin{minipage}[c]{246pt}
\def\arraystretch{1.5}
\def\tabcolsep{8pt}
(a) Ideal (000$\bar{1}$)--O surface:\hfill\strut\\[4pt]
\begin{tabular}{lccc}
Site:           & `on--top' & `hcp--hollow' & `fcc--hollow' \\ \hline
$\Delta E$ [eV] &   +3.16   &     0.0       &    +0.05
\end{tabular}\\
\vspace{12pt}
(b) Hydrogen covered (000$\bar{1}$)--O surface:\hfill\strut\\[4pt]
\begin{tabular}{lccc}
Site:           & `on--top' & `hcp--hollow' & `fcc--hollow' \\ \hline
$\Delta E$ [eV] &   +1.78   &     0.0       &  $-$0.02
\end{tabular}
\end{minipage}
\end{center}
\caption{\label{tab:h}
Relative stability of the high-symmetry adsorption sites for (a) the
surface O atoms of the ideal O--terminated surface and (b) the OH--groups
of the H saturated surface for a full monolayer coverage. The $\Delta E$ are
the calculated energy differences per O--atom/OH--group for moving the topmost
O/OH--surface layer from the regular lattice position of the wurtzite
structure (`hcp--hollow--site') to the `on--top' and `fcc--hollow' position.}
\end{table}

\begin{table}[b]
\noindent
\begin{center}
\begin{minipage}[c]{246pt}
\def\arraystretch{1.5}
\def\tabcolsep{5pt}
\begin{tabular}{lcccccc}
$c_{\rm H}$   &  1   &  3/4    &  2/3    &  1/2    &  1/3    &  1/4\\ \hline
$E_{\rm b}$ [eV] & $-$0.71 & $-$1.10 & $-$1.25 & $-$1.90 & $-$2.12 & $-$2.20
\end{tabular}
\end{minipage}
\end{center}
\caption{\label{tab:hcov}
Calculated binding energies $E_{\rm b}$ per H atom for dissociating H$_2$
molecules and forming hydrogen layers of coverage $c_{\rm H}$.}
\end{table}

\subsection{Hydrogen Adsorption}
\label{sec:hcov}

We turn now to a situation in which the (000$\bar{1}$)--O surface is in
equilibrium with a H$_2$ gas phase. H$_2$ molecules can dissociate, and
hydrogen atoms may adsorb at the surface thereby forming OH-groups with
the surface oxygen ions. Before we start to calculate the surface energy
for different surface models with H adatoms, we consider the possibility
that the preferred adsorption site of these OH-groups is no longer the
regular lattice site of the O ions. Three different high-symmetry adsorption
sites are conceivable above the underlying Zn layer: an `on-top' position,
a `hcp-hollow site', which is the regular lattice position for the O ions
in the wurtzite structure, and a `fcc-hollow site'.

First we consider the clean surface without adsorbed hydrogen. Then
we see from Table~\ref{tab:h} that indeed the `hcp-hollow' sites are the
most stable positions for the O surface layer. However, moving the whole
layer to `fcc-hollow' sites costs only a small amount of energy. Turning
to the hydroxylated surface with a full monolayer coverage of hydrogen
we find that the OH-groups now prefer the `fcc-hollow site'. So by gradually
adding hydrogen, the regular lattice site of the surface O ions becomes
unstable relative to the `fcc-hollow site'. However, the energy difference
between `fcc-' and `hcp-hollow sites' is so small that we have neglected this
effect in all further calculations and have only considered `hcp-hollow sites'
for surface oxygen atoms and OH--groups.

\begin{figure}[!t]
\noindent
\epsfxsize=246pt
\epsffile{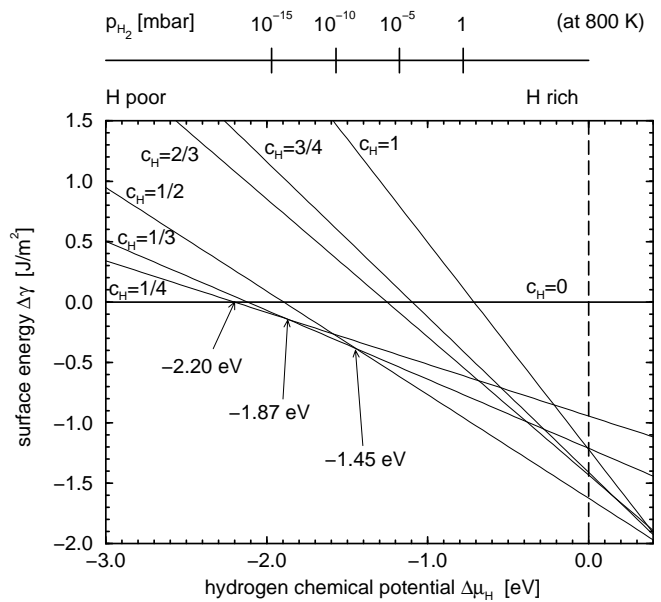}
\caption{\label{fig:hcov}
Surface free energy $\Delta\gamma$ of the polar O-terminated (000$\bar{1}$)--O
surfaces with different coverages of hydrogen $c_{\rm H}$ as function of the
hydrogen chemical potential $\Delta\mu_{\rm H}$. In the top $x$--axis, the
chemical potential $\Delta\mu_{\rm H}(T,p)$ has been translated into a
pressure scale for the fixed temperature of $T$=800\,K. The vertical dash line
indicates the upper bound for $\Delta\mu_{\rm H}$.}
\end{figure}

In the next step we construct different surface models of hydrogen covered
(000$\bar{1}$)--O surfaces using a similar procedure as in Sec.~\ref{sec:ovac}.
We take slabs with (1$\times$1), (1$\times$2), (1$\times$3) and (2$\times$2)
surface unit cells, and we add different amounts of hydrogen to create
hydrogen coverages of $c_{\rm H}$ of 1/4, 1/3, 1/2, 2/3, 3/4 and 1 monolayer.
The calculated surface energy changes $\Delta\gamma$ relative to the
clean (000$\bar{1}$)--O surface are plotted in Fig.~\ref{fig:hcov} as a
function of the hydrogen chemical potential $\Delta\mu_{\rm H}$. A very
similar behavior as in Section~\ref{sec:ovac} arises: At H--rich conditions
the structure with a 1/2 monolayer hydrogen coverage is the most stable
surface, in agreement with mechanism (III). On the other hand, at H--poor
conditions the clean surface without hydrogen becomes the most stable
structure. As a new feature we find that between the two limiting cases the
surfaces with 1/3 and 1/4 monolayer coverage are slightly more stable for
small intervals of the chemical potentials.\cite{comment3}  Thus, by lowering 
the chemical potential $\Delta\mu_{\rm H}$ from H--rich to H--poor conditions
it is possible to gradually reduce the hydrogen coverage from 1/2 monolayer to
1/3, 1/4 and zero coverage. Translating the chemical potential into
temperature and pressure conditions we find that we will start to remove H at
UHV-pressures at a temperature of roughly 750\,K, which is in reasonable
agreement with the experimental observation.\cite{woell}  The other surface
structures with hydrogen coverages larger than 1/2 are always higher in energy
and will be unstable for all temperatures and hydrogen partial pressures. In
particular a surface with a full monolayer of hydrogen as postulated from the
results of the HAS experiment\cite{woell} is not likely to exist in
thermodynamic equilibrium. However, from the intensity of the He--specular
peak it was deduced that only about 0.1\,\% of the (000$\bar{1}$)--O surface
consists of flat terraces with diameters exceeding 50\,{\AA} which contribute
to the (1$\times$1) HAS signal.\cite{woell}  Therefore, the H covered surface
with a (1$\times$1) HAS diffraction pattern may well be a minority phase which
is formed under suitable kinetic conditions.

In Table~\ref{tab:hcov} we have summarized the H binding energies $E_{\rm b}$
per atom which we have calculated as energy difference between the ideal
(000$\bar{1}$)--O surface plus $1/2\,E_{\rm H_2}$ and surface structures with
H coverages of $c_{\rm H}$. We see that it becomes rapidly unfavorable to
adsorb more hydrogen as soon as the concentration for ideal charge
compensation of the polar surface of 1/2 monolayer is reached. The reason
for this behavior is the same as in the case of the oxygen vacancies: Up to
1/2 monolayer of hydrogen we fill the partially occupied O--$2p$--band, beyond
1/2 monolayer the O--$2p$--band is completely filled and we have to populate
the conduction band. The decrease in $E_{\rm b}$ from 1/2 to 1/4 monolayer
coverage indicates that also at low coverages a weak repulsive interaction
between the hydrogen atoms remains. This is the reason why also the
low-coverage structures appear in the surface phase diagram. If we extrapolate
$E_{\rm b}$ toward zero coverage of the surface we can estimate an initial
binding energy of about 2.3\,eV per H atom for the dissociative adsorption of
H$_2$.

\subsection{Water Dissociation}
\label{sec:water}

As evident from Fig.~\ref{fig:hcov} hydrogen adsorption plays a major role
for the stabilization of the polar (000$\bar{1}$)--O surface and for almost
all conceivable experimental conditions hydrogen will be present on the
surface. But until now we have only considered molecular hydrogen as the
reservoir for the surface hydrogen. However, in many chemical reactions and
catalytic processes also water is present. Therefore we will briefly explore
if also water can act as a source for surface hydrogen
%DM .
in the presence of the (0001)--Zn face. 

In a DFT study using the hybrid B3LYP functional Wander and
Harrison\cite{wander_h} found that dissociating water and forming a full
H and OH monolayer on the O-- and Zn--terminated surface, respectively,
is energetically unfavorable by 0.1\,eV. As adsorption sites for the
OH--groups they assumed the Zn `on-top' positions which would be the
next lattice sites for O ions if the crystal is extended. However,
on the polar surfaces two more high-symmetry adsorption sites exist:
the `hcp-hollow site' position above atoms in the second surface layer
and a `fcc-hollow' site with no atoms beneath.

Using the same adsorption geometry as Wander and Harrison we also find that
dissociating water is energetically unfavorable with a slightly larger energy
cost of 0.3\,eV. However, as shown in Table~\ref{tab:oh}, the configuration
with the OH groups adsorbed at the `fcc-hollow sites' instead of the
`on-top' positions is much lower in energy. Considering the correct adsorption
positions for the OH groups we now find that even for the thermodynamically
unstable monolayer coverages the dissociation of water is energetically
preferred by about 0.4\,eV. Taking only 1/2 monolayer coverages into account,
this energy gain will be significantly larger.

\begin{table}[!t]
\noindent
\begin{center}
\begin{minipage}[c]{246pt}
\def\arraystretch{1.5}
\def\tabcolsep{8pt}
\begin{tabular}{lccc}
Site:           & `on--top' & `hcp--hollow' & `fcc--hollow' \\ \hline
$\Delta E$ [eV] &    0.0    &  $-$0.04      &  $-$0.72
\end{tabular}
\end{minipage}
\end{center}
\caption{\label{tab:oh}
Relative stability of the different OH adsorption sites on the polar
Zn--terminated surface, calculated for a monolayer coverage.}
\end{table}

\subsection{Surface Phase Diagram}
\label{sec:phase}

Finally we combine the results of the previous subsections and assume that
the polar (000$\bar{1}$)--O surface is now simultaneously in equilibrium with
an O$_2$ and a H$_2$ gas phase. In addition to the surface models described
in Sec.~\ref{sec:ovac} and Sec.~\ref{sec:hcov} we have furthermore considered
various mixed structures of O vacancies and adsorbed H atoms in the
(1$\times$2), (1$\times$3) and (2$\times$2) surface unit cells.

\begin{figure}[!t]
\noindent
\epsfxsize=246pt
\epsffile{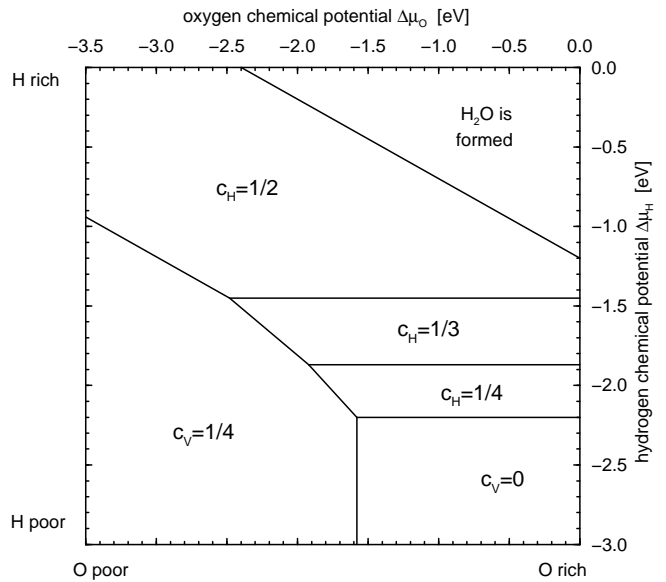}
\caption{\label{fig:phase}
Phase diagram of the polar O-terminated (000$\bar{1}$) surface in equilibrium
with H and O particle reservoirs controlling the chemical potentials
$\Delta\mu_{\rm H}$ and $\Delta\mu_{\rm O}$, based on selected superstructures
as explained in the text. The lowest-energy surface structures are labeled by
the concentrations of oxygen vacancies $c_{\rm V}$ and hydrogen adatoms
$c_{\rm H}$. The upper right area indicates conditions under which H$_2$O
condensates on the surface.}
\end{figure}

The surface free energy now depends on two chemical potentials
$\Delta\mu_{\rm O}$ and $\Delta\mu_{\rm H}$. The graphs of
Fig.~\ref{fig:ovac} and \ref{fig:hcov} therefore have to be extended to a
3-dimensional plot. Such a diagram would be rather complex and hard to follow.
%DM  up. 
The most important information contained in the plot of the surface free
energies is which of the surface models has the lowest surface energy for a
given combination of chemical potentials $\Delta\mu_{\rm O}$ and
$\Delta\mu_{\rm H}$. This information is better visualized if we project the
3-dimensional diagram onto the ($\Delta\mu_{\rm O}$,$\Delta\mu_{\rm H}$) plane
and only mark the regions for which a certain surface structure is the most
stable one. The result is a phase diagram of the (000$\bar{1}$)--O surface
which is shown in Fig.~\ref{fig:phase}.

The surface phase diagram in Fig.~\ref{fig:phase} summarizes in a condensed
fashion the essential results of our study. This phase diagram is dominated
by two structures: a surface with 1/2 monolayer of adsorbed H and a
hydrogen-free surface with 1/4 of the oxygen atoms removed. These are the two
scenarios denoted as mechanism (II) and (III) in the Introduction, indicating
that filling the O--$2p$--bands is a very important mechanism to stabilize the
(000$\bar{1}$)--O surface. Next to these two phases we find two structures in
which H is gradually removed from the surface and only at very H--poor
conditions and when plenty of oxygen is available, the ideal O--terminated
surface stabilized by mechanism (I) becomes the most stable structure.

None of the additionally considered mixed structures which simultaneously
contain O vacancies and H adatoms appears in the phase diagram. This is not
very surprising since for the given sizes of the surface unit cells no
combination of O vacancies and H adatoms exists that leads to fully occupied
O--$2p$--bands. However, it is very well conceivable that for larger surface
unit cells mixed structures with, for example, an O vacancy concentration
of 1/8 and a H coverage of 1/4, become more stable which would then appear
as new phases between the H covered and the O vacancy structures.

Relating the chemical potentials to temperature conditions and partial
pressures of the gas phase shows, see Fig.~\ref{fig:phase}, that for almost
all realistic experimental conditions hydrogen will be present at the
(000$\bar{1}$)--O surface. Even at UHV-conditions with a low hydrogen
partial pressure one has to go to rather high temperatures to fully remove
all hydrogen. In this case, a surface structure with O vacancies will become
the most stable one. In order to stabilize the ideal O--terminated surface an
oxygen atmosphere with an extremely low content of hydrogen (and also water
vapor) is necessary, which is basically not achievable in experiment.

In Table~\ref{tab:dij} we have summarized the surface relaxations for the
three most important surface structures appearing in the surface phase
diagram, Fig.~\ref{fig:phase}. For the extended surface structures with
H adatoms and O vacancies we have averaged in each atomic plane parallel to
the surface the atomic displacements, and we define the interlayer distances
$d$ as the separation of the averaged atomic positions. Depending on the
surface structure and the charge compensation process, very different surface
relaxations occur. The largest relaxations are found for the clean,
defect-free surface termination with a compression of the first double-layer
distance of almost 50\,\% and also a significant contraction of the second and
subsequent double-layer spacings. This is in agreement with other previous
ab-initio studies.\cite{noguera,zno34,wander,carlsson}  After filling the
partially occupied surface bands by adsorbing 1/2 monolayer of H or by
removing 1/4 of the O ions, however, the surface layers relax back to almost
truncated-bulk-like positions. Thus, for surfaces with a lower H adatom or
O vacancy concentration, surface relaxations between the two extremes are
conceivable. This may explain why experimentally very different results for
the surface relaxations were found. GIXD measurements\cite{wander,jedrecy}
have predicted an inward relaxation of the upper O--layer with a contraction
of the first double-layer distance of 40\,\% and 20\,\%, respectively, whereas
from LEED\cite{duke} and LEIS\cite{overbury} experiments it was concluded that
the first double-layer spacing is close to its bulk value.

\begin{table}[!t]
\noindent
\begin{center}
\begin{minipage}[c]{246pt}
\def\arraystretch{1.2}
\def\tabcolsep{3pt}
\begin{tabular}{llll}
\strut\quad $d$ & ideal surface & H covered & O vacancies \\ \hline
H$-$O$_1$ & & 0.1825 \\
O$_1$$-$Zn$_2$ &
              0.0628 ($-$48\,\%) & 0.1207    (0.0\,\%) & 0.1151 ($-$4.6\,\%)\\
Zn$_2$$-$O$_3$ &
              0.3985  (+5.1\,\%) & 0.3779 ($-$0.4\,\%) & 0.3767 ($-$0.7\,\%)\\
O$_3$$-$Zn$_4$ &
              0.1077 ($-$11\,\%) & 0.1246   (+3.2\,\%) & 0.1238   (+2.6\,\%)\\
Zn$_4$$-$O$_5$ &
              0.3813  (+0.5\,\%) & 0.3773 ($-$0.5\,\%) & 0.3772 ($-$0.6\,\%)
\end{tabular}
\end{minipage}
\end{center}
\caption{\label{tab:dij}
Summary of the averaged interlayer distances $d$ (given in fractions of the
theoretical bulk lattice constant $c$) and the relative changes with respect
to the bulk values (in parentheses) for three different surface structures:
the ideal, defect-free (000$\bar{1}$)--O surface, the H covered surface
with 1/2 monolayer of hydrogen and the surface with a vacancy concentration
of 1/4. The subscripts refer to surface layers numbered from the surface
plane. The theoretical PBE bulk values for the interlayer distances are
$d_{\rm O-Zn}$=0.1208\,$c$ (in bilayer) and $d_{\rm Zn-O}$=0.3792\,$c$
(between bilayers) and the lattice constant was calculated to be
$c$=5.291\,{\AA} (see Ref.~\protect\onlinecite{zno34}).}
\end{table}

%------------------------------------------------------------------------------

\section{Summary and Conclusions}
\label{sec:summary}

By combining first-principles density-functional calculations with a
thermodynamic formalism we have determined lowest-energy structures of the
polar O--terminated (000$\bar{1}$)--O surface of ZnO in thermal equilibrium
with an O$_2$ and H$_2$ gas phase. This scheme allows us to extend our
zero-temperature and zero-pressure DFT results to more realistic temperature
and pressure conditions which are usually applied in surface science
experiments or in catalytic processes like the methanol synthesis, and thus
to bridge computationally the `pressure gap'.

The essential result of this approach is a phase diagram of the
(000$\bar{1}$)--O surface which is plotted in Fig.~\ref{fig:phase}.
From this surface phase diagram we predict that hydrogen is adsorbed at the
(000$\bar{1}$)--O surface for a wide range of temperatures and H$_2$ partial
pressures, including UHV--conditions. This is in agreement with the recent
observations of a HAS experiment\cite{woell} and was also confirmed in a
study of the CO adsorption on the polar ZnO surfaces.\cite{zno35} 

We find a H binding energy of roughly 2.3\,eV per atom if molecular hydrogen
dissociates and adsorbs at the clean O--terminated surface. Furthermore
we predict that in situations where both polar surface terminations are
present (for example for powder samples) also the dissociative adsorption of
water with H and OH--groups being adsorbed at the O-- and Zn--terminated
surface, respectively, is energetically preferable. However, as soon as
a coverage of 1/2 monolayer of hydrogen is reached, the energy gain of
adsorbing more hydrogen on the (000$\bar{1}$)--O surface drops very rapidly
with increasing hydrogen coverage. Therefore no stable phases with more than
1/2 monolayer H coverage appear in the surface phase diagram,
Fig.~\ref{fig:phase}. In particular, a structure with a full monolayer
of H as predicted in Ref.~\onlinecite{woell} is not very likely to exist
globally in thermodynamic equilibrium (which does not exclude a kinetic
or local stabilization).

Going to low hydrogen partial pressures and higher temperatures it is
possible to gradually remove the hydrogen from the surface and to form
stable phases with less than 1/2 monolayer coverage of hydrogen. If all
hydrogen is removed, oxygen vacancies will be created as was speculated in
Ref.~\onlinecite{woell}. However, we find that a surface with a vacancy
concentration of 1/4 is much more stable than a missing-row structure where
1/3 of the oxygens has been removed. Therefore, based on the limited set of
surface structures taken into consideration in our study, we currently do not
understand the (1$\times$3) structure observed in Ref.~\onlinecite{woell}.

Finally, at higher oxygen partial pressures the O vacancies will be filled
and the clean, defect-free O--terminated surface becomes the most stable
structure. However, the hydrogen partial pressure has to be very low so
that we consider it very unlikely that a clean, defect-free (000$\bar{1}$)--O
surface can be observed in experiment.

%------------------------------------------------------------------------------

\section{Acknowledgments}

The author would like to thank Dominik Marx, Christof W\"oll, Georg Kresse,
and Ulrike Diebold for fruitful discussions. The work was supported by SFB~558
and FCI.

%------------------------------------------------------------------------------

%------------------------------------------------------------------------------


\begin{thebibliography}{999}
\bibitem{comment1} For simplicity we use an ionic model to introduce the
unfamiliar reader into the stability problem of polar surfaces. However, the
same consequences for the atomic and electronic structure of polar surfaces
arise if a pure covalent bonding as in the electron-counting/auto-compensation
model or a mixed ionic/covalent bonding as in the bond-transfer model are
assumed.\cite{noguera}  A more general approach was given by M.W. Finnis
in Ref.~\onlinecite{finnis1}, where the polarity of a surface is related to
the excess of an atomic species and the non-stoichiometry of a surface,
independently of any model of the chemical bond.
\bibitem{comment2} In a covalent model one may argue as follows: At the polar
ZnO surfaces one of the 4 nearest-neighbor bonds of the surface atoms is
broken. Since O contributes 6 and Zn 2 electrons to the 4 covalent bonds,
partially occupied dangling bonds will appear at the surfaces with 3/2 and
1/2 electrons per surface atom at the O-- and Zn--terminated surface,
respectively, forming a 3/4 and a 1/4 filled surface band. This is often
referred to as ``charge neutrality condition of the surface''. The partially
filled bands can be avoided by a reconstruction of the surface, loss of
surface atoms or by adsorption of charged species. For a more general
description see Ref.~\onlinecite{noguera} and \onlinecite{finnis1}.
\bibitem{comment3} It might also be possible that we can gradually increase
the O--vacancy concentration from zero to 1/4. However, in our calculations
we did not consider surface structures with vacancy concentrations less than
1/4 so that we can not make a definite statement.
\end{thebibliography}
\end{document}